\documentclass[aps,preprint,superscriptaddress]{revtex4}
\usepackage{graphicx}
\usepackage{amsmath}
\usepackage{dcolumn}
\usepackage{color}
\usepackage{epstopdf}
\usepackage[caption=false]{subfig}

\def\V{\textbf{V}}
\def\v{\textbf{v}}
\def\x{\textbf{x}}
\def\b{\textbf{b}}

\def\f{\textbf{f}}

\def\B{\textbf{B}}

\def\e{\textbf{e}}
\def\pa{\partial}

\begin{document}

\title{Optimized boundary driven flows for dynamos in a sphere}

\author{I. V. Khalzov}
\affiliation{University of Wisconsin-Madison, 1150 University Avenue, Madison, Wisconsin 53706, USA}
\affiliation{Center for Magnetic Self Organization in Laboratory and Astrophysical Plasmas}
\author{B. P. Brown}
\affiliation{University of Wisconsin-Madison, 1150 University Avenue, Madison, Wisconsin 53706, USA}
\affiliation{Center for Magnetic Self Organization in Laboratory and Astrophysical Plasmas}
\author{C. M. Cooper}
\affiliation{University of Wisconsin-Madison, 1150 University Avenue, Madison, Wisconsin 53706, USA}
\affiliation{Center for Magnetic Self Organization in Laboratory and Astrophysical Plasmas}
\author{D. B. Weisberg}
\affiliation{University of Wisconsin-Madison, 1150 University Avenue, Madison, Wisconsin 53706, USA}
\affiliation{Center for Magnetic Self Organization in Laboratory and Astrophysical Plasmas}
\author{C. B. Forest}
\affiliation{University of Wisconsin-Madison, 1150 University Avenue, Madison, Wisconsin 53706, USA}
\affiliation{Center for Magnetic Self Organization in Laboratory and Astrophysical Plasmas}

\date{\today}

\begin{abstract}
We perform numerical optimization of the axisymmetric flows in a sphere to minimize the critical magnetic Reynolds number $Rm_{cr}$ required for dynamo onset.  The optimization is done for the class of laminar incompressible flows of von K\'arm\'an type satisfying the steady-state Navier-Stokes equation. Such flows are determined by equatorially antisymmetric profiles of driving azimuthal (toroidal) velocity specified at the spherical boundary. The model is relevant to the Madison plasma dynamo experiment (MPDX), whose spherical boundary is capable of differential driving of plasma in the azimuthal direction. We show that the dynamo onset in this system depends strongly on details of the driving velocity profile and the fluid Reynolds number $Re$. It is found that the overall lowest $Rm_{cr}\approx200$ is achieved at $Re\approx240$ for the flow, which is hydrodynamically marginally stable. We also show that the optimized flows can sustain dynamos only in the range $Rm_{cr}<Rm<Rm_{cr2}$, where $Rm_{cr2}$ is the second critical magnetic Reynolds number, above which the dynamo is quenched. Samples of the optimized flows and the corresponding dynamo fields are presented.    
\end{abstract}

\maketitle

\section{Introduction}

The creation of specific flows of an electrically conducting fluid is the key element and major challenge in experimental investigation of dynamo phenomenon. Appropriate candidates for this role are the flows capable of the dynamo action, a number of them are theoretically studied in literature \cite{Roberts_1972, Ponomarenko_1973, Kumar_1975, Dudley_1989, Gubbins_2000a, Gubbins_2000b, TCF_num_2000, VKS_num_2003, Moss_2008, Spence_2009}. In most of these studies kinematic dynamos are considered, i.e., the magnetic induction equation is treated as an eigenvalue problem for an unknown magnetic field and a prescribed laminar flow. The model flow is usually chosen in a simple analytical form, which is not necessarily determined from the fluid dynamics. In fact, the majority of analyzed flows leading to kinematic dynamos do not satisfy the Navier-Stokes equation (e.g., flows from Refs.~\cite{Roberts_1972, Kumar_1975, Dudley_1989,Gubbins_2000a, Gubbins_2000b, Moss_2008}), so their structure cannot be reproduced exactly in the laboratory. Nevertheless, it is possible to obtain experimentally a flow sufficiently close to the model one, which may result in the dynamo action. 

In the past decade, several groups constructed dynamo experiments intended to achieve such flows with liquid metals \cite{Gailitis_2000, Peffley_2000, Stieglitz_2001, Forest_2002, Bourgoin_2002, Monchaux_2007, Lathrop_2011}. Although  the obtained flows were highly turbulent in all the experiments, their mean parts were expected to sustain a dynamo field. However, only three experiments were successful in dynamo demonstration: experiments in Riga, Latvia \cite{Gailitis_2000}, and Karlsruhe, Germany \cite{Stieglitz_2001}, where the flows were strongly constrained and the influence of turbulence was small, and a von K\'arm\'an sodium experiment in Cadarache, France, where ferromagnetic impellers played the critical role \cite{Monchaux_2007, Verhille_2010}. It seems likely that hydrodynamic turbulence in unconstrained flows significantly inhibits dynamo onset.

The presence of turbulence is an inevitable problem in all liquid metal dynamo experiments. This is due to the extremely low magnetic Prandtl numbers of the liquid metals, i.e., the ratio of kinetic viscosity $\nu$ to resistivity $\eta$ or, equivalently, the ratio of magnetic Reynolds number to fluid Reynolds number $Pm\equiv\nu/\eta=Rm/Re$ (e.g., for liquid sodium $Pm\sim10^{-5}$). In order to reach  the magnetic Reynolds numbers sufficient for dynamo excitation ($Rm\sim10^1-10^2$), very high fluid Reynolds numbers ($Re\sim10^6-10^7$) are required. As a result, the corresponding flows in experiments are always turbulent, making it difficult to achieve dynamos and to compare experimental data with predictions of laminar kinematic theory. Note that the liquid metal laboratory dynamos with $Pm\ll1$ are in the regime of  the solar convection zone and the interiors of planets, while hot accretion disks and galaxies have  $Pm\gg1$.

\begin{figure}[bt]
\centering
\includegraphics[scale=1]{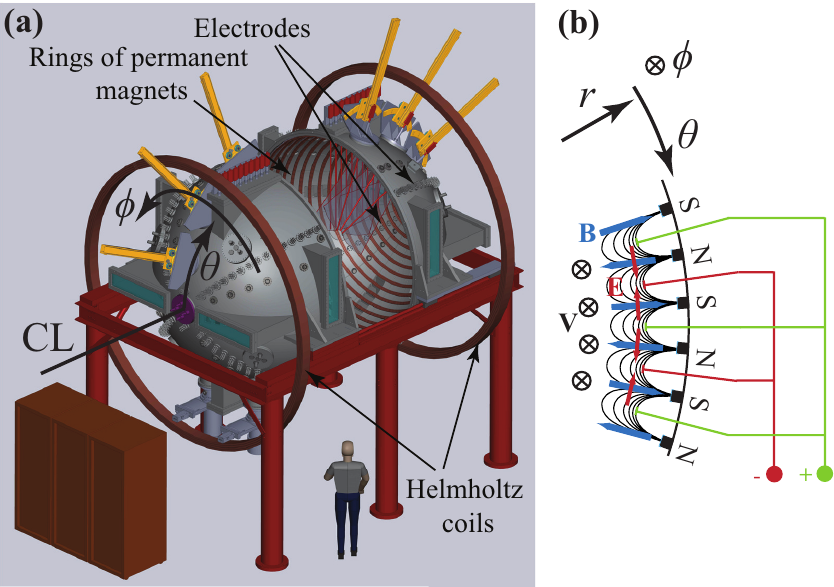}
\caption{The Madison plasma dynamo experiment (MPDX): (a) a sketch of the experiment; (b) the electrode configuration near the wall for driving plasma velocity $v_\phi(\theta)$. The spherical system of coordinates $(r,\theta,\phi)$ is shown. Center line  (CL) corresponds to the axis of symmetry. Reprinted with permission from Phys.~Plasmas \textbf{19}, 104501 (2012) \cite{Khalzov_2012b}. Copyright 2012 American Institute of Physics.}\label{MPDX}
\end{figure}

The present study is motivated by construction of the Madison plasma dynamo experiment (MPDX, Fig.~\ref{MPDX}), which is designed to investigate dynamos excited by controllable flows of plasma \cite{Forest_2008, Spence_2009}. The use of plasma as the electrically conducting fluid gives experimentalists flexibility in choosing a regime of dynamo operation. By adjusting experimental controls, one can change driving velocity, density, electron and ion temperatures, etc. This makes it possible to adjust $Pm$, $Rm$ and $Re$ at will and study laminar dynamos with $Pm\sim1$ and $Rm\sim Re\sim10^2$ (Table \ref{t1}). Such flexibility is advantageous over the liquid metal dynamo experiments. The experimental vessel in MPDX is an aluminum sphere of 3 meter in diameter [Fig.~\ref{MPDX}(a)]. Plasma is confined by an axisymmetric multicusp magnetic field created by 36 equally spaced rings of permanent magnets with alternating polarity. The plasma filling the vessel is mostly unmagnetized since the multicusp field is localized near the vessel wall. The novel feature of the experiment is the mechanism for creating controllable plasma flows [Fig.~\ref{MPDX}(b)]. An electric field applied across the multicusp magnetic field drives the edge of the plasma azimuthally, while viscosity couples momentum from the edge to the unmagnetized core. Nearly arbitrary profiles of plasma azimuthal (toroidal) velocity $v_\phi(\theta)$ can be imposed at the sphere's boundary by modulating the electric field as a function of polar angle $\theta$ using discrete electrodes. This concept of plasma stirring has been successfully tested in the plasma Couette experiment (PCX) \cite{Collins_2012}, and it allows a unique way to conduct laboratory studies of various astrophysical  phenomena including the dynamo \cite{Spence_2009, Khalzov_2011}, the magnetorotational instability \cite{Ebrahimi_2011}, and the Parker instability \cite{Khalzov_2012}.        

\begin{table}[btp]
\caption{Expected parameters of MPDX. Dimensionless numbers $Re$, $Rm$ and $Pm$  are estimated from the Braginskii equations \cite{Brag_1965} (see corresponding formulas in Refs.~\cite{Spence_2009, Khalzov_2011}). }
\centering
\begin{tabular}{lccc}
\hline
\hline
Quantity & Symbol & Value & Unit \\
\hline 
Radius of sphere & $R_0$ & 1.5 & m \\
Peak driving velocity  & $V_0$ & $0-20$ & km/s\\
Average number density &   $n_0$  &  $10^{17}-10^{19}$ & m$^{-3}$\\
Electron temperature & $T_e$ & $2-10$ & eV \\
Ion temperature & $T_i$ &  $0.5-4$ & eV \\
Ion species &  & H, He, Ne, Ar &\\
Ion mass & $\mu_i$ & 1, 4, 20, 40 & amu\\
\hline
Fluid Reynolds  & $Re$ & $0-10^5$ & \\
Magnetic Reynolds & $Rm$ & $0-2\times10^3$ & \\
Magnetic Prandtl & $Pm$ & $10^{-3}-5\times10^3$ & \\
\hline
\end{tabular}
\label{t1}
\end{table}

As shown in Refs.~\cite{Elsasser_1946, Bullard_1954}, the toroidal motion alone does not sustain a dynamo magnetic field, some poloidal flow is necessary. In a bounded sphere the poloidal flow develops self-consistently from differential toroidal rotation, which is controlled in MPDX by boundary driving velocity $v_\phi(\theta)$. It can be shown using the Navier-Stokes equation that the intensity of poloidal flow is determined by fluid Reynolds number $Re$:  the poloidal flow is normally stronger for the larger values of $Re$. There is a minimum amount of poloidal motion necessary for dynamo action \cite{Busse_1975}. From this point of view, the large values of fluid Reynolds number $Re$ are desirable in the experiment.  At the same time, to avoid turbulent regime of the flow, the values of $Re$ should be less than the hydrodynamic instability threshold. We emphasize here again that fluid Reynolds number $Re$ in the unmagnetized MPDX plasma can be varied by changing ion parameters of plasma: density $n_0$, temperature $T_i$ and mass $\mu_i$.   

The poloidal motion is required for dynamo excitation, but does not guarantee it. The dynamo onset is very sensitive to the details of flow structure and corresponding profile of driving azimuthal velocity $v_\phi(\theta)$. The ability to create arbitrary profiles of $v_\phi(\theta)$ in MPDX raises the question of their optimization for the dynamo excitation. The goal of our study is to find numerically the optimized profiles of driving velocity $v_\phi(\theta)$ and the corresponding equilibrium flow structures in the model relevant to MPDX. We refer to a flow as optimized if it minimizes the critical magnetic Reynolds number $Rm_{cr}$ required for the kinematic dynamo onset. Physically, a lower $Rm_{cr}$ means a lower driving velocity and/or lower electron temperature (higher resistivity) of the plasma needed for achieving dynamo. 

The problem of flow optimization for the kinematic dynamo in spherical geometry was addressed by many researchers \cite{Love_1996a, Love_1996b, OConnell_2001, Holme_2003, Gubbins_2008}. The necessity of balancing the relative amplitudes of the toroidal and poloidal flow components for the dynamo action was originally noticed for the Kumar and Roberts flow (KR flow) \cite{Kumar_1975} and later for the Dudley and James flows (DJ flows) \cite{Dudley_1989}. More rigorously, the idea of flow optimization was introduced by Love and Gubbins \cite{Love_1996a, Love_1996b}, who optimized the relative amplitudes of the four spherical harmonic components of the KR flow, keeping their radial dependences fixed. O'Connell with co-authors \cite{OConnell_2001} and Holme \cite{Holme_2003} optimized DJ flows, allowing the radial structure of spherical harmonic components to vary.  Gubbins and co-authors showed that the optimized flow of KR type maximizes the non-axisymmetric part of kinetic helicity (see Ref.~\cite{Gubbins_2008} and references therein).   

It appears that all flow optimizations considered in the dynamo literature so far are performed  for simplified model flows of a particular type (either KR or DJ flows). The structure of these flows is usually prescribed by their type and is not determined self-consistently from the hydrodynamic equations. In this sense, the optimized flows found are not realistic and cannot be reproduced in an experiment.  In contrast to these studies, in the present paper we deal only with realistic flows found as solutions to the steady-state Navier-Stokes equation with boundary conditions specified by the driving velocity profiles $v_\phi(\theta)$.  

The structure of the paper is as follows. In Sec.~II, we briefly describe the model of MPDX used in our study (the basic equations of the model and methods of their solution are given in Appendix A). In Sec.~III, we discuss the factors that influence the dynamo threshold. In Sec.~IV, the results of the flow optimization are reported. In Sec.~V, we summarize our main findings.

\section{Model and optimization procedure}

We perform our study in the framework of single-fluid magnetohydrodynamics (MHD), which is a good approximation for the MPDX plasma. For simplicity, we ignore the effects of plasma compressibility and the details of plasma confinement and drive near the wall. We neglect the multicusp magnetic field and the applied electric field and assume that the velocity profile is specified at the sphere's  boundary. As shown in Ref.~\cite{Khalzov_2011} for the model relevant to PCX (a cylindrical prototype for MPDX), these ignored details only play a role in the relatively thin boundary layers, and are not essential in the bulk of unmagnetized plasma.  

The equations of our model in non-dimensional form are
\begin{eqnarray}
\label{mhd_NS} 0&=&\nabla^2\v-Re\big[(\v\cdot\nabla)\v+\nabla p\big],\\
\label{mhd_v} \gamma_v\tilde\v&=&\nabla^2\tilde\v-Re\big[(\tilde\v\cdot\nabla)\v+(\v\cdot\nabla)\tilde\v+\nabla\tilde p\big],\\
\label{mhd_b} \gamma_b\B&=&\nabla^2\B+Rm\nabla\times(\v\times\B),\\ 
\label{mhd_div} 0&=&\nabla\cdot\v=\nabla\cdot\tilde\v=\nabla\cdot\B, 
\end{eqnarray}
where $\v=\V/V_0$ and $p=P/(\rho_0 V_0^2)$ are normalized velocity and plasma pressure in equilibrium, $\tilde\v$ and $\tilde p$ are their perturbations near equilibrium, respectively. Two dimensionless numbers, fluid Reynolds $Re$ and magnetic Reynolds $Rm$ are defined as 
$$
Re=\frac{R_0V_0}{\nu},~~~Rm=\frac{R_0V_0}{\eta}.
$$     
In defining the normalized quantities, we use the peak driving velocity $V_0$, radius of the sphere $R_0$ (a unit of length throughout the paper), plasma mass density  $\rho_0$, the kinematic viscosity $\nu$ and the magnetic diffusivity $\eta$ (all three assumed to be constant and uniform).  Eq.~(\ref{mhd_NS}) is the Navier-Stokes equation describing the equilibrium velocity $\v$. Since we are interested only in a linear (kinematic) stage of dynamo, we do not include the Lorentz force due to the dynamo field in Eq.~(\ref{mhd_NS}). Eq.~(\ref{mhd_v}) is the Navier-Stokes equation linearized near the equilibrium velocity $\v$. It constitutes an eigenvalue problem for the velocity perturbation $\tilde\v$ and its growth rate $\gamma_v$. By solving Eq.~(\ref{mhd_v}), one can establish the hydrodynamic stability properties of the equilibrium velocity $\v$. Eq.~(\ref{mhd_b}) is the kinematic dynamo problem for the unknown dynamo magnetic field $\B$ and its growth rate $\gamma_b$.  Note that the growth rates in Eqs.~(\ref{mhd_v}) and (\ref{mhd_b}) are normalized by the corresponding inverse diffusion times: $\gamma_v$ is given in units of $\nu/R_0^2$ and $\gamma_b$ is given in units of $\eta/R_0^2$. Normalization of magnetic field $\B$ is arbitrary due to linearity of Eq.~(\ref{mhd_b}) with respect to $\B$.         

We restrict our study to the axisymmetric equilibrium flows only, which have dependences of the form $\v(r,\theta)$ in the spherical system of coordinates $(r,\theta,\phi)$. Here $r$ is the normalized radius ($0\leq r\leq1$), $\theta$ is the polar angle ($0\leq\theta\leq\pi$) and $\phi$ is the azimuthal or toroidal angle ($0\leq\phi\leq2\pi$). Exploiting the geometry of the problem, we expand the divergence-free fields $\v$, $\tilde\v$ and $\B$ in a spherical harmonic basis \cite{Bullard_1954} and substitute these expansions into Eqs.~(\ref{mhd_NS})-(\ref{mhd_b}). The resulting equations and methods of their numerical solution are given in Appendix \ref{app}.  
 
Eqs.~(\ref{mhd_NS})-(\ref{mhd_b}) are supplemented by appropriate boundary conditions. In our model, the boundary condition for equilibrium velocity $\v$ is specified by the driving velocity profile at the sphere's wall:
\begin{equation}\label{BCv}
\v\big|_{r=1}=v_\phi(\theta)\e_\phi,~~~0\leq\theta\leq\pi,
\end{equation}
where $v_\phi(\theta)$ is a function of the polar angle $\theta$ with physical restriction $v_\phi(0)=v_\phi(\pi)=0$.  In the present  study we consider only the flows of von K\'arm\'an type, i.e., the flows, whose driving azimuthal velocity $v_\phi(\theta)$ is antisymmetric with respect to equator ($\theta=\pi/2$). In such flows, the dynamo growth rate $\gamma_b$ is purely real near the dynamo onset, which is shown in Sec.~IV. As a result, the critical magnetic Reynolds number $Rm_{cr}$ can be easily found  by setting $\gamma_b=0$ in Eq.~(\ref{mhd_b}). For numerical convenience, we use the Fourier expansion of $v_\phi(\theta)$:
\begin{equation}\label{vb_VK}
v_\phi(\theta)=\sum\limits_{\substack{n=2 \\ \textrm{even}~n}}^{N}a_n\sin n\theta=a_2\bigg(\sin 2\theta+\beta_4\sin 4\theta+\beta_6\sin 6\theta+\dots\bigg),~~~\max\limits_{0\leq\theta\leq\pi} v_\phi(\theta)=1.
\end{equation}
Here we keep only even harmonics of $\theta$ due to equatorial antisymmetry of function $v_\phi(\theta)$, and introduce parameters $\beta_n\equiv a_n/a_2$. In addition, by adjusting coefficient $a_2$ we normalize $v_\phi(\theta)$ so that its maximum is 1. With this definition of driving velocity $v_\phi(\theta)$, the resulting equilibrium flow (as well as corresponding $Rm_{cr}$) is uniquely determined by a set of independent parameters $\beta_n$ and fluid Reynolds number $Re$. 

The boundary condition for velocity perturbation $\tilde\v$ in case of impenetrable, no-slip wall is 
\begin{equation}\label{BCdv}
\tilde\v\big|_{r=1}=0.
\end{equation}

Specification of boundary conditions for magnetic field $\B$ plays an essential role in dynamo studies. As shown in Ref.~\cite{Khalzov_2012b} for the model relevant to MPDX, the critical magnetic Reynolds number $Rm_{cr}$ is sensitive to the wall magnetic permeability but not affected by the wall resistivity. In our present study, we assume the non-ferritic insulating wall, so that the normal component of the electric current is zero at the boundary, and the normal component of the magnetic field matches the external vacuum solution. These conditions, along with Eqs.~(\ref{BCv}) and (\ref{BCdv}), can be conveniently represented in terms of a spherical harmonic basis. The corresponding  equations are given in Appendix \ref{app}. 

We briefly describe the optimization procedure used in the study.  First, we solve Eq.~(\ref{mhd_NS}) to find the axisymmetric equilibrium velocity field $\v$ for a given set of  driving parameters $\beta_n$ and fluid Reynolds number $Re$.  Then, we solve Eq.~(\ref{mhd_v}) using $\v$ to find the eigenvalue with the largest real part $\textrm{Re}\{\gamma_v\}$.  Depending on this eigenvalue, the equilibrium flow  is classified as stable ($\textrm{Re}\{\gamma_v\}<0$), marginally stable ($\textrm{Re}\{\gamma_v\}=0$), or unstable ($\textrm{Re}\{\gamma_v\}>0$). The hydrodynamically unstable flows are not examined in our kinematic dynamo study, since they will develop into non-axisymmetric structures, which contradicts our analysis. We note that the instability of these flows is usually due to the modes with azimuthal numbers $m=1$ or $m=2$. In addition, the growth rate $\gamma_v$ is always real near the instability threshold. 

As the third step, we consider kinematic dynamo problem given by Eq.~(\ref{mhd_b}) with velocity $\v$. Our calculations show that the fastest growing (or least decaying) dynamo modes have azimuthal mode number  $m=1$ (generally corresponding to equatorial dipoles), so we restrict our consideration to these modes only. In addition, for the equilibrium flows of von K\'arm\'an type (with equatorially antisymmetric azimuthal velocity) the dynamo growth rate $\gamma_b$ is always real near the dynamo instability threshold, so when $Rm=Rm_{cr}$ the growth rate is zero $\gamma_b=0$. This circumstance allows us to significantly  simplify the procedure of finding $Rm_{cr}$. We solve Eq.~(\ref{mhd_b}) with $\gamma_b=0$ as a generalized eigenvalue problem for $Rm_{cr}$. The minimal positive eigenvalue found corresponds to the required $Rm_{cr}$. If there are no positive numbers among calculated eigenvalues, then the dynamo cannot be excited or it is excited with $\textrm{Im}\{\gamma_b\}\ne0$. 

These three steps define $Rm_{cr}$ implicitly as a function of fluid Reynolds number $Re$ and independent driving parameters $\beta_n$. For a given $Re$ we search for a minimum of $Rm_{cr}$ in a multi-dimensional space of the driving parameters $\beta_n$ and determine  the optimized flow. Multi-dimensional numerical minimization in our study is realized via the downhill simplex method (also known as Nelder-Mead or amoeba method \cite{Press_2007}). 

\section{Dynamo onset}\label{Sec3}

\begin{figure}[tbp]
\centering
\includegraphics[scale=1]{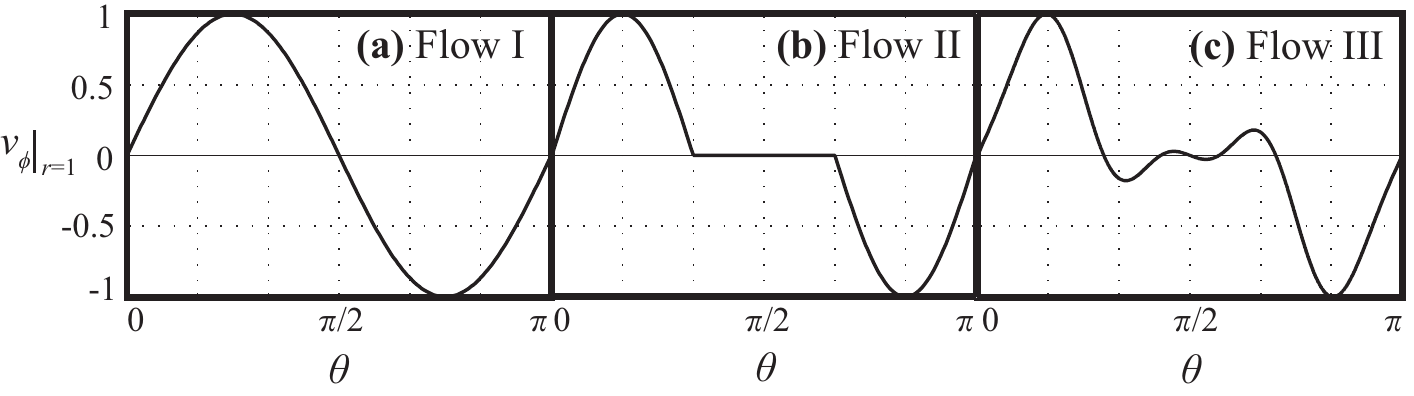}
\caption{Profiles of driving boundary velocity $v_\phi(\theta)$ for flows I, II and III.}\label{Flows_bc}
\end{figure}  

\begin{figure}[tbp]
\centering
\includegraphics[scale=1]{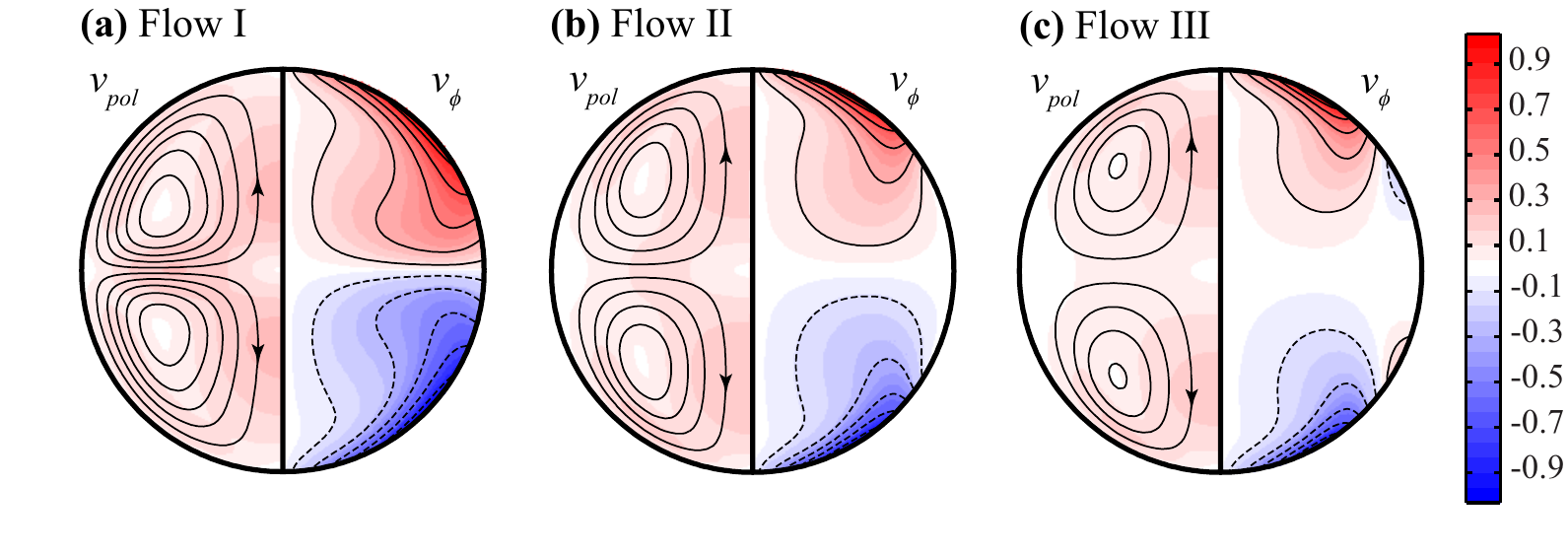}
\caption{Equilibrium structures of flows I, II and III for fluid Reynolds number $Re=150$. The left half of each figure shows stream lines of poloidal velocity $v_{pol}$ superimposed on its absolute values depicted in colors, the right half shows a contour plot of azimuthal velocity $v_\phi$ (dashed curves denote values of $v_\phi<0$). The vertical central lines represent the axis of symmetry. }\label{Flows}
\end{figure}

\begin{figure}[tbp]
\centering
\includegraphics[scale=1]{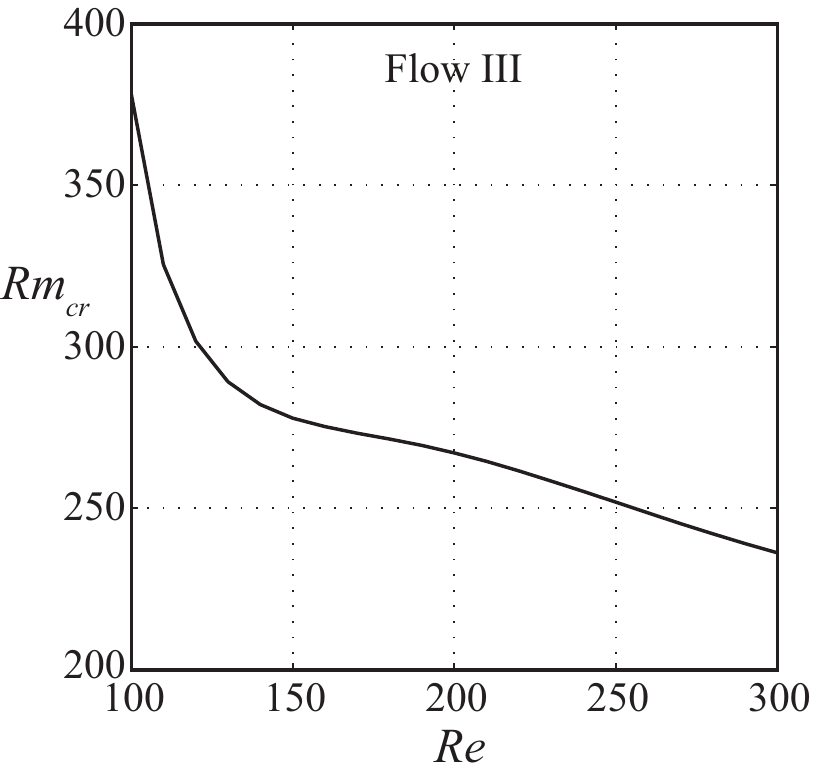}
\caption{Critical magnetic Reynolds $Rm_{cr}$ as a function of fluid Reynolds $Re$ for flow III.}\label{Rm_Re_III}
\end{figure}

In this section we study the effects that influence the dynamo onset. For this purpose we consider three axisymmetric equilibrium flows (denoted as flow I, II and III, respectively) driven by different profiles of boundary azimuthal velocity $v_\phi(\theta)$ as shown in Fig.~\ref{Flows_bc}. The corresponding equilibrium flow structures for $Re=150$ are presented in Fig.~\ref{Flows}. Flow I is driven by boundary velocity $v_\phi(\theta)=\sin2\theta$, i.e., only the first harmonic in Eq.~(\ref{vb_VK}) is kept. Driving velocity profile of flow II consists of  half-sines with opposite signs in the intervals $0<\theta<\pi/3$  and $2\pi/3<\theta<\pi$. Boundary velocity of flow III is taken from Ref.~\cite{Spence_2009}; it is similar to that of flow II, but has several reversals (changes of sign) in equatorial region. The driving Fourier coefficients for  these flows are given in Table \ref{t2}. The coefficients are truncated at $n=20$, this is the number of harmonics used in the kinematic dynamo calculations. We note that among these flows only flow III results in dynamo action (dependence of critical magnetic Reynolds number $Rm_{cr}$ on fluid Reynolds number $Re$ is given in Fig.~\ref{Rm_Re_III}). The reasons for this are analyzed below.    

\begin{table}[tbp]
\caption{Fourier coefficients $a_n$ of driving velocities $v_\phi(\theta)$ for flows I, II and III.}
\centering
\begin{tabular}{cccc}
\hline
\hline
~~$n$~~ & ~~Flow I~~ & ~~Flow II~~ & ~~Flow III~~ \\
\hline 
2 & 1 & ~0.6616 & ~0.4853 \\
4 & 0 & ~0.4726 & ~0.5235 \\
6 & 0 & 0 & ~0.0467 \\
8 & 0 & -0.0601 & -0.1516 \\
10 & 0 & ~0.0364 & 0 \\
12 & 0 & 0 & 0 \\
14 & 0 & -0.0177 & 0 \\
16 & 0 & ~0.0134 & 0 \\
18 & 0 & 0 & 0 \\
20 & 0 & -0.0085 & 0 \\
\hline
\end{tabular}
\label{t2}
\end{table}

\begin{figure}[tbp]
\centering
\includegraphics[scale=1]{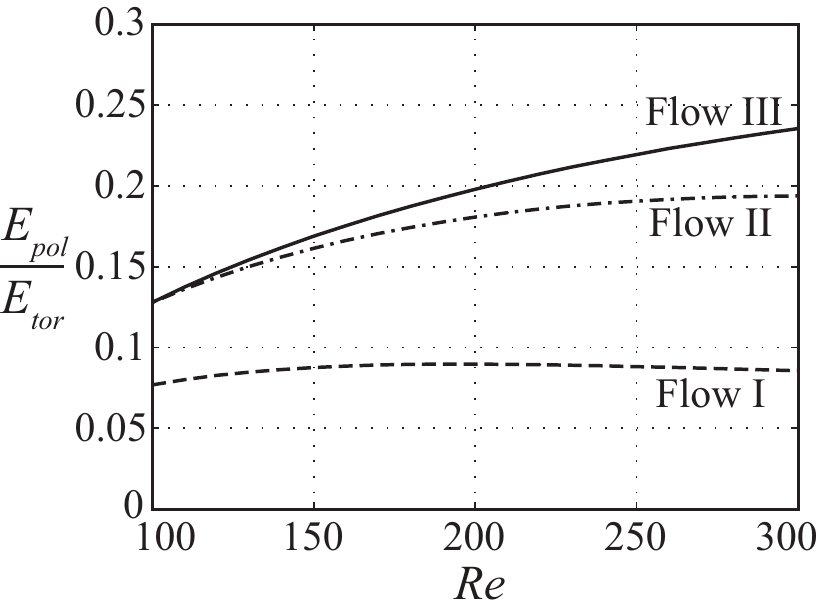}
\caption{Ratio of poloidal to toroidal kinetic energies $E_{pol}/E_{tor}$ as a function of fluid Reynolds number $Re$ for flows I, II and III.}\label{Epol}
\end{figure}

As established in Refs. \cite{Elsasser_1946, Bullard_1954}, the pure toroidal flow cannot excite a dynamo in a spherical geometry, a finite poloidal component of flow is necessary too. Therefore, dynamo onset should depend on the ratio of poloidal to toroidal flow amplitudes or, equivalently, the ratio of poloidal to toroidal kinetic energies. For the counter-rotating axisymmetric flow of Dudley and James  \cite{Dudley_1989} it was found that this ratio must be in a certain range in order to excite a dynamo.  In a boundary driven flow, this ratio is determined by the details of the driving velocity profile and value of fluid Reynolds number $Re$. Fig.~\ref{Epol} shows the ratio of poloidal to toroidal kinetic energies $E_{pol}/E_{tor}$ as a function of $Re$ for flows I, II and III.  The figure suggests that the failure of flow I to support a dynamo most likely results from insufficient relative intensity of its poloidal component. 

Fig.~\ref{Epol} is obtained assuming that flows I, II and III are axisymmetric at $100<Re<300$. However, in reality their axial symmetry breaks when they become  hydrodynamically unstable. The corresponding threshold values of fluid Reynolds number, above which the non-axisymmetric instabilities develop in these flows, are $Re_{I}\approx115$ (instability with azimuthal mode number $m=2$), $Re_{II}\approx207$ ($m=2$) and $Re_{III}\approx305$ ($m=1$). 

The presence of poloidal flow component with large enough amplitude compared to toroidal component is a necessary but not sufficient condition for dynamo onset. The similar driving velocity profiles of flows II and III produce similar ratios of poloidal to toroidal kinetic energies $E_{pol}/E_{tor}$ (Fig.~\ref{Epol}). However, they are completely different from the dynamo point of view: flow III leads to a dynamo action, while flow II does not. Such difference can be explained by the details of driving velocity profiles. Namely, the presence of reversals in $v_\phi(\theta)$ in equatorial region of flow III appears to be crucial for the dynamo onset. All the optimized flows found in Sec.~\ref{Sec4} possess this property. Currently, the role of these reversals for dynamo onset is not understood completely.

\section{Results of flow optimization}\label{Sec4}

\begin{figure}[tbp]
\centering
\includegraphics[scale=1]{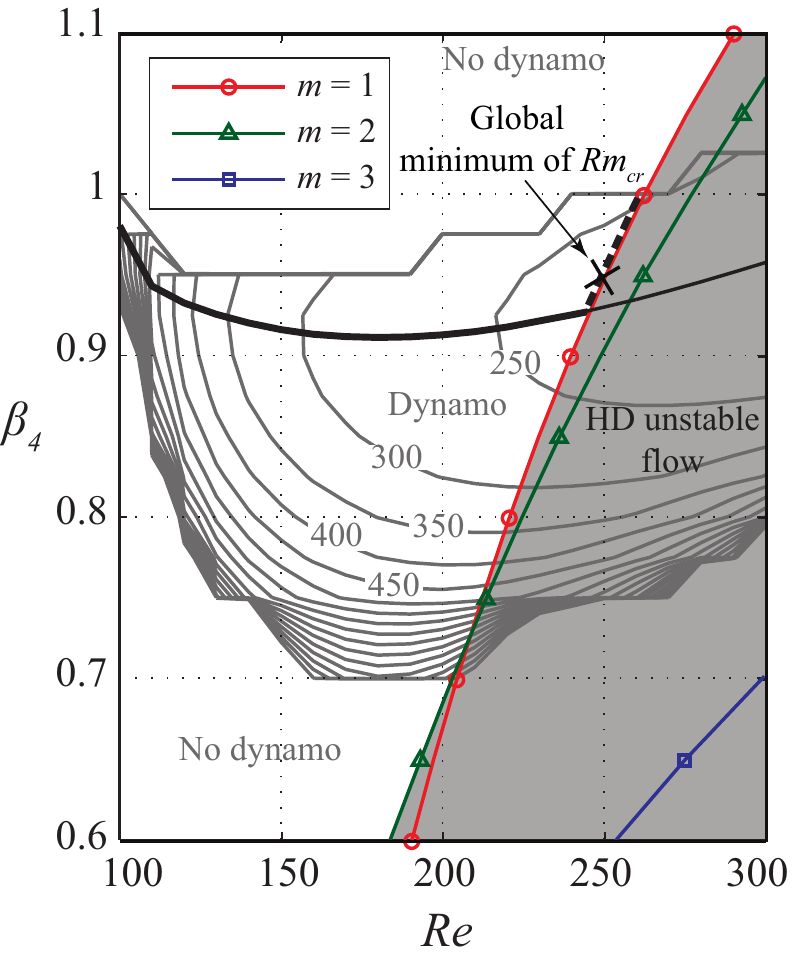}
\caption{Contour plot of critical magnetic Reynolds number $Rm_{cr}$ as a function of fluid Reynolds number $Re$ and driving parameter $\beta_4\equiv a_4/a_2$. Contours of $Rm_{cr}$ are shown. The shaded area denotes  the hydrodynamically unstable region, with stability boundaries shown for azimuthal modes $m=1,2,3$ (curves labeled with symbols). Each point of the solid black curve corresponds to the optimized stable flow that minimizes $Rm_{cr}$ at a given value of $Re$. Points of the dashed black curve correspond to the optimized flows at the boundary of hydrodynamic stability. Symbol ``$\times$" denotes  the point ($Rm_{cr}\approx237$, $Re\approx250$), at which the global minimum of $Rm_{cr}$ is achieved. The segmentation of the dynamo/no dynamo boundary is due to discrete scan of the plane $Re-\beta_4$.}\label{a4_Re}
\end{figure}

In this section we report the results of the flow optimization, which is performed to minimize the critical magnetic Reynolds number $Rm_{cr}$ required for the dynamo onset. First, we consider equilibrium flows driven by $v_\phi(\theta)$ with  two lowest Fourier harmonics in Eq.~(\ref{vb_VK}): 
$$
v_\phi(\theta)=a_2\bigg(\sin 2\theta+\beta_4\sin 4\theta\bigg),~~~\max\limits_{0\leq\theta\leq\pi} v_\phi(\theta)=1.
$$
Such equilibrium flows are uniquely determined by two independent parameters: Fourier  coefficients ratio $\beta_4\equiv a_4/a_2$  and fluid Reynolds number $Re$. The corresponding $Rm_{cr}$ is also a function of these two parameters. The contour plot of this function $Rm_{cr}(Re, \beta_4)$ is shown in Fig.~\ref{a4_Re}.  Dynamo action exists only in a bounded domain of the parameter space, approximately in the range $100<Re<270$ and $0.7<\beta_4<1$. Note that if $\beta_4=0$, then $v_\phi(\theta)\to\sin 2\theta$, and the dynamo action is not possible (flow I from Sec.~\ref{Sec3}). Increase in fluid Reynolds number $Re$ makes the background equilibrium flow hydrodynamically unstable with respect to non-axisymmetric modes (shaded area), and our kinematic dynamo analysis is not valid in this case. Scanning $Re$ we determine values of $\beta_4$, which lead to optimized equilibrium flows minimizing $Rm_{cr}$; these values form a solid black curve in Fig.~\ref{a4_Re}. For $Re>245$ the optimized flows are at the boundary of hydrodynamic stability (dashed curve). The global minimum of the critical magnetic Reynolds number in this case is $Rm_{cr}\approx237$ achieved at  $Re\approx250$.

\begin{figure}[tbp]
\centering
\includegraphics[scale=1]{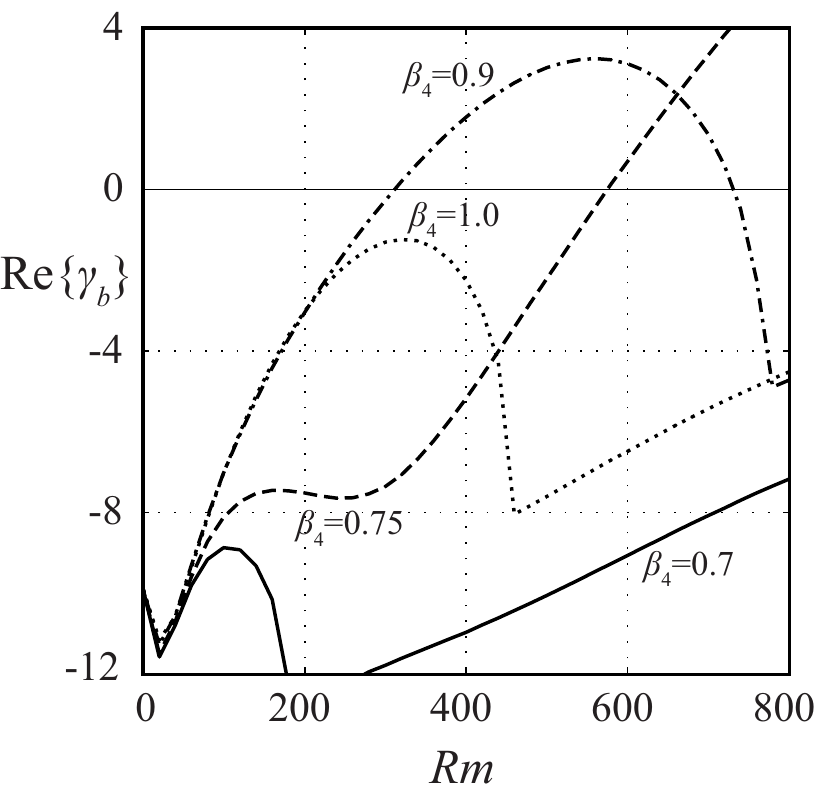}
\caption{Real part of dynamo growth rate $\textrm{Re}\{\gamma_b\}$ as a function of magnetic Reynolds number $Rm$ for fluid Reynolds number $Re=150$ and different values of driving parameters $\beta_4=0.7$ (solid curve), $\beta_4=0.75$ (dashed curve), $\beta_4=0.9$ (dashed-dotted curve) and $\beta_4=1.0$ (dotted curve).}\label{gams}
\end{figure}

In Fig.~\ref{a4_Re},  the lower  ($\beta_4\approx0.7$) and the upper ($\beta_4\approx1$)  dynamo/no dynamo  transitions  have different behavior of $Rm_{cr}$. This difference is clarified in Fig.~\ref{gams}, which  shows curves of $\gamma_b(Rm)$  at $Re=150$ for several values of $\beta_4$.   $Rm_{cr}$ increases as $\beta_4$ decreases from  0.75 to 0.7. This corresponds roughly to tilting the curve $\gamma_b(Rm)$ to the right in Fig.~\ref{gams}. When $\beta_4$ approaches $0.7$ from above, $Rm_{cr}$ goes to infinity, and there is no dynamo for $\beta_4<0.7$ (the lower dynamo/no dynamo transition). Increasing $\beta_4$ from  $0.9$ to $1.0$    corresponds roughly  to shifting the parabolic curve $\gamma_b(Rm)$  down,  below the $\gamma_b=0$ line  in Fig.~\ref{gams}. This explains why the upper dynamo/no dynamo transition at $\beta_4\approx1$  occurs at finite values of $Rm_{cr}$.

\begin{figure}[tbp]
\centering
\includegraphics[scale=0.9]{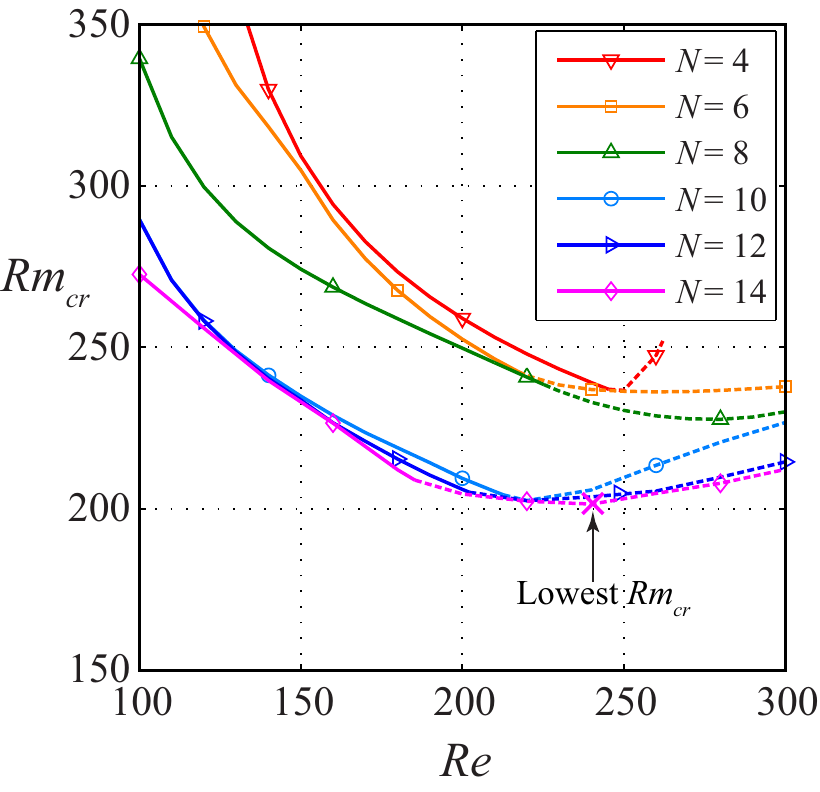}
\caption{Critical magnetic Reynolds $Rm_{cr}$ as a function of fluid Reynolds $Re$ for optimized flows with different number of driving harmonics. Solid curves correspond to the optimized hydrodynamically stable flows, dashed curves correspond to the optimized flows at the boundary of hydrodynamic stability. $N$ denotes the number of the highest non-zero Fourier harmonic in the driving velocity $v_\phi(\theta)$. Symbol ``$\times$" denotes the overall lowest value of $Rm_{cr}\approx200$, which is achieved at $Re\approx240$.}\label{Rm_Re}
\end{figure}

\begin{figure}[btp]
\centering
\includegraphics[scale=1]{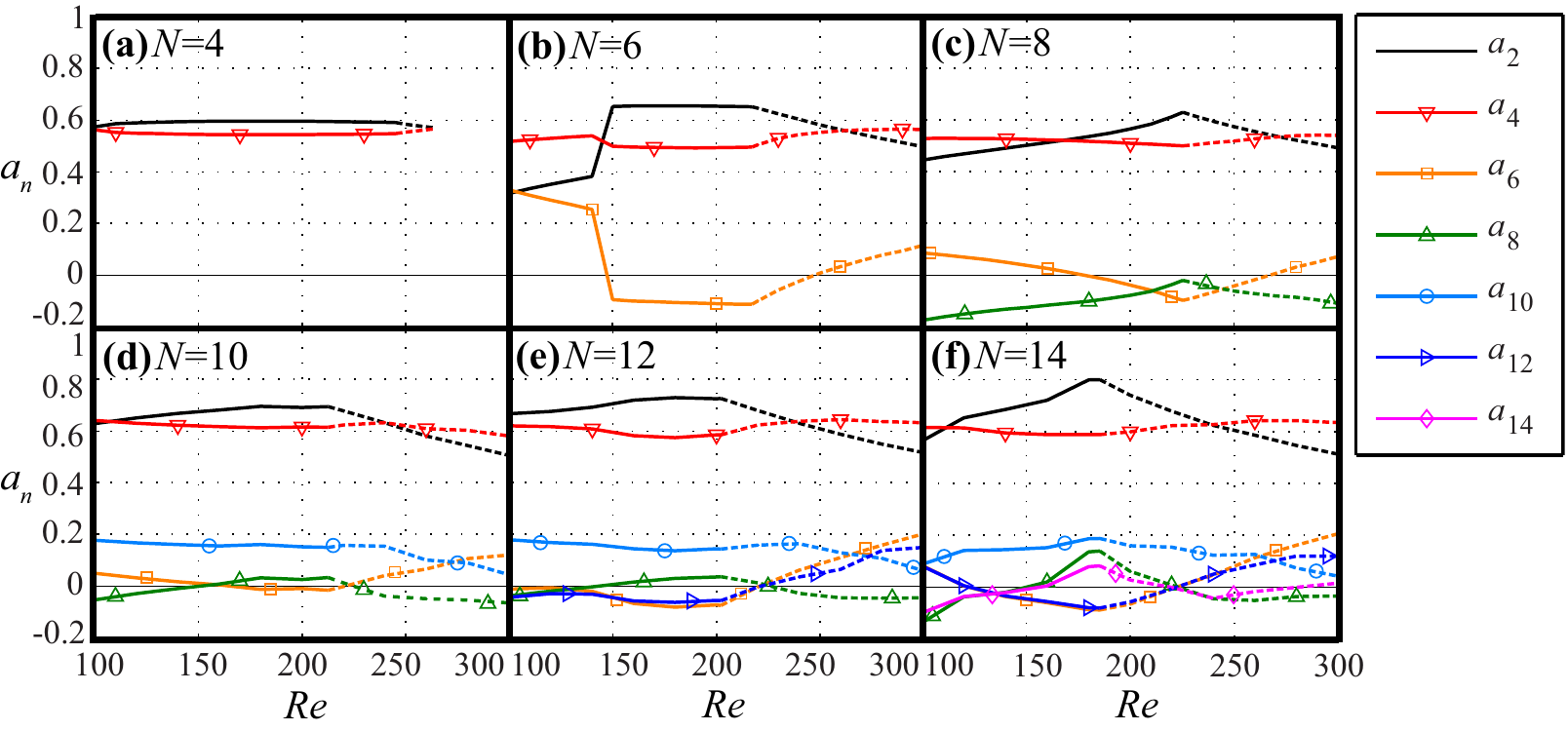}
\caption{Dependences of driving Fourier coefficients $a_n$  on fluid Reynolds $Re$ in optimized flows with different number of driving harmonics. Values of $Re$ are scanned with step $\Delta Re=10$.  As in Fig.~\ref{Rm_Re}, solid curves correspond to the stable flows, dashed curves correspond to the marginally stable flows.}\label{a_tot}
\end{figure}

We perform similar analysis for the flows with different number of Fourier harmonics in driving boundary velocity $v_\phi(\theta)$. The respective cases are marked according to the number of the highest non-zero harmonic in Eq.~(\ref{vb_VK}) (for example, $N=8$ means the flows driven by all even harmonics up to $\sin 8\theta$). The results of the analysis are summarized in Figs.~\ref{Rm_Re}  and \ref{a_tot} for six cases with values of $N$ ranging from $N=4$ to $N=14$. The curves in Fig.~\ref{Rm_Re} show the lowest possible $Rm_{cr}$ achievable for a given $Re$ by optimizing the flows with a different number of driving harmonics. In all cases, $Rm_{cr}$ decreases with increasing $Re$ for the optimized hydrodynamically stable flows (solid curves) and reaches a minimum for the flows at the stability boundary (dashed curves). The overall lowest $Rm_{cr}\approx200$ is obtained at  $Re\approx240$ for $N=14$.  Fig.~\ref{a_tot} shows the dependences of driving Fourier coefficients $a_n$ on fluid Reynolds number $Re$ in the optimized flows. Note that only two coefficients $a_2$ and $a_4$ are relatively large in all cases ($a_4\sim a_6\sim0.6$), the magnitude of others is normally less than 0.2. 

\begin{figure}[tbp]
\centering
\includegraphics[scale=1]{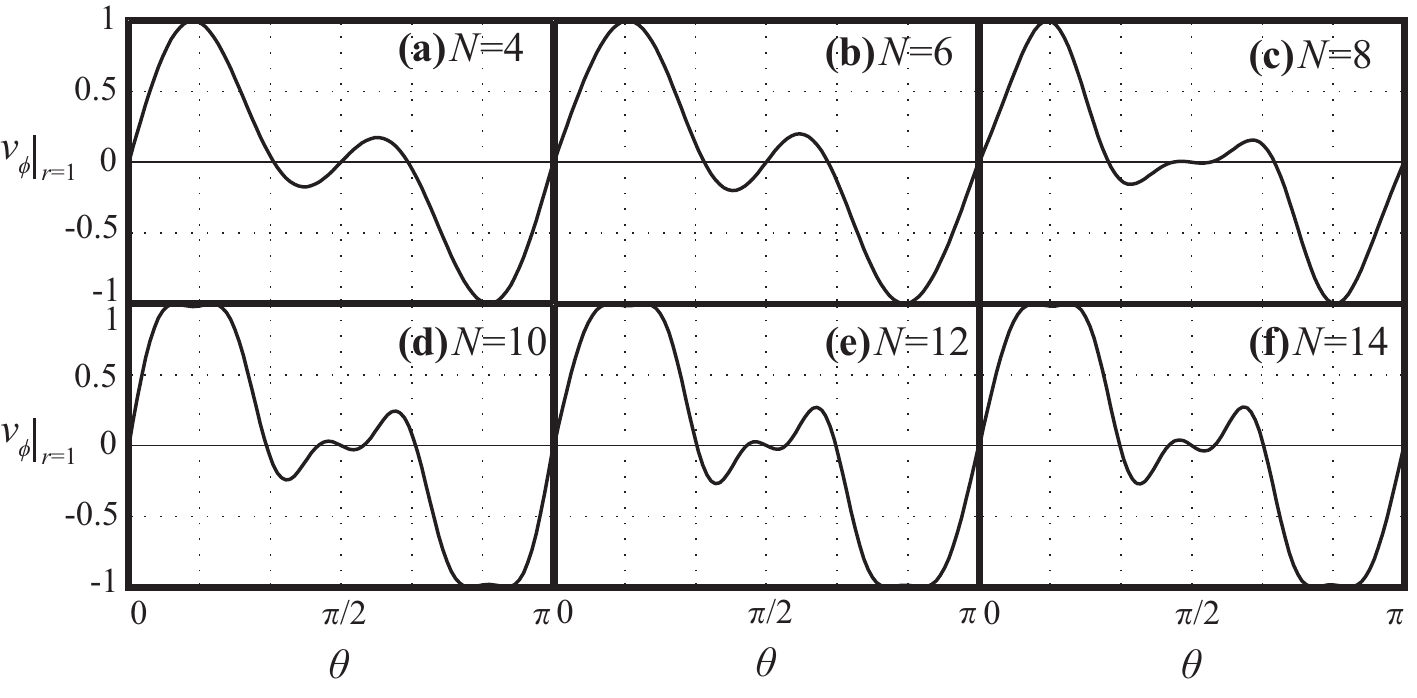}
\caption{Profiles of optimized azimuthal velocity $v_\phi(\theta)$ at the boundary for different number of driving harmonics, fluid Reynolds number is $Re=150$.}\label{vb_tot}
\end{figure}  
\begin{figure}[tbp]
\centering
\includegraphics[scale=1]{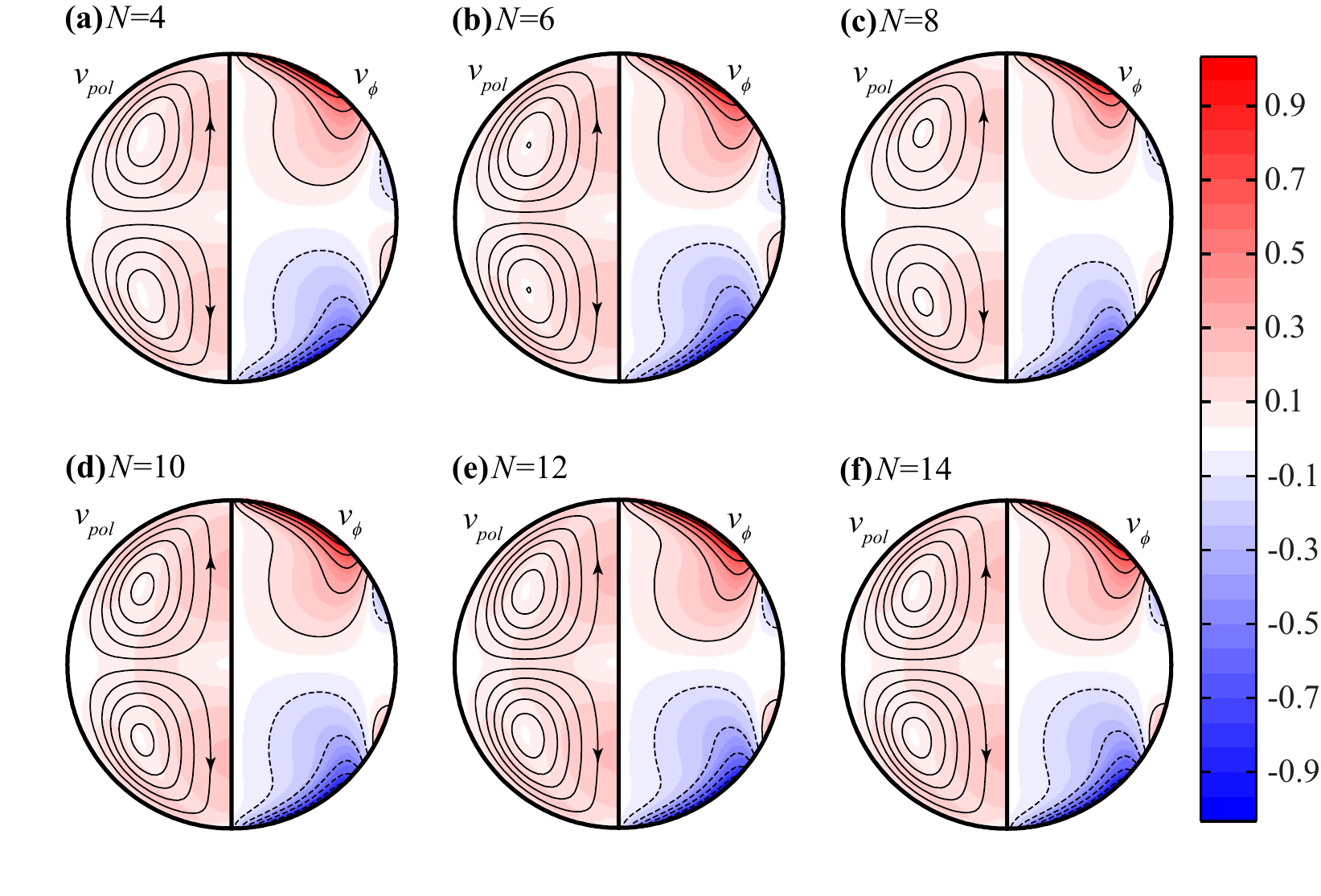}
\caption{Axisymmetric equilibrium flows corresponding to optimized driving velocities from Fig. \ref{vb_tot} at fluid Reynolds number $Re=150$. Notations are the same as in Fig.~\ref{Flows}.}\label{v_tot}
\end{figure}

Fig.~\ref{vb_tot} demonstrates the samples of optimized profiles of boundary velocity $v_\phi(\theta)$ for different number of driving harmonics (optimization is done at $Re=150$). The corresponding structures of equilibrium flows are shown in Fig.~\ref{v_tot}. The optimized driving velocities $v_\phi(\theta)$ exhibit changes of sign in equatorial region. As mentioned in Sec.~\ref{Sec3}, the presence of such reversals appears to be crucial for the dynamo action. Also we note that for the cases with $N=10$, $N=12$ and $N=14$, both the optimized profiles of $v_\phi(\theta)$ and the corresponding flow structures are very similar. This indicates that at this stage, the optimized flows are not strongly affected by higher driving harmonics. 

\begin{figure}[tbp]
\centering
\includegraphics[scale=1]{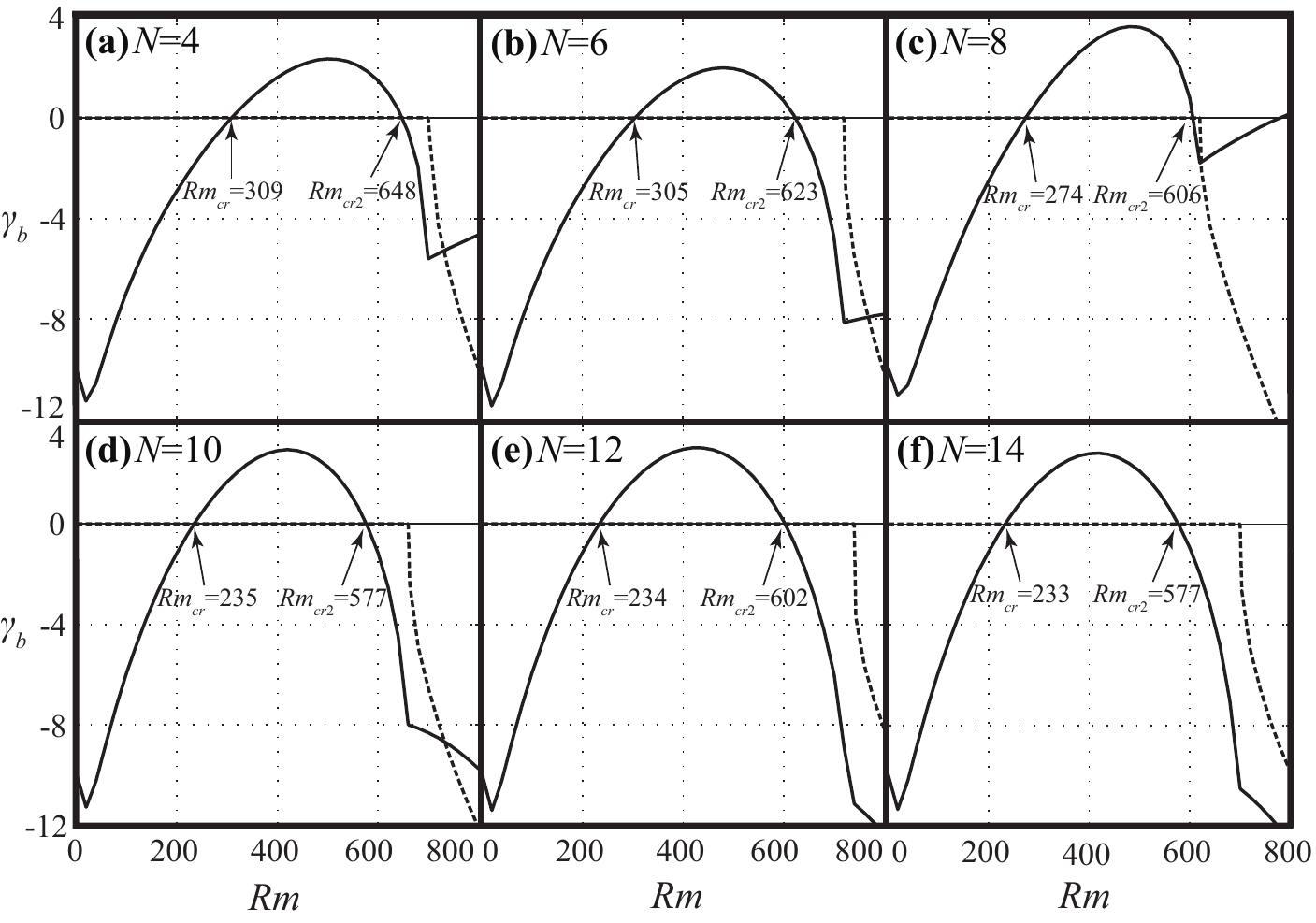}
\caption{Dependences of real (solid curves) and imaginary (dashed curves) parts of dynamo growth rate $\gamma_b$ on magnetic Reynolds number $Rm$ for the optimized  flows shown in Fig.~\ref{v_tot}. Calculations are done for the fastest dynamo mode (with azimuthal mode number $m=1$). }\label{gam_tot}
\end{figure}

\begin{figure}[tbp]
\centering
\includegraphics[scale=1]{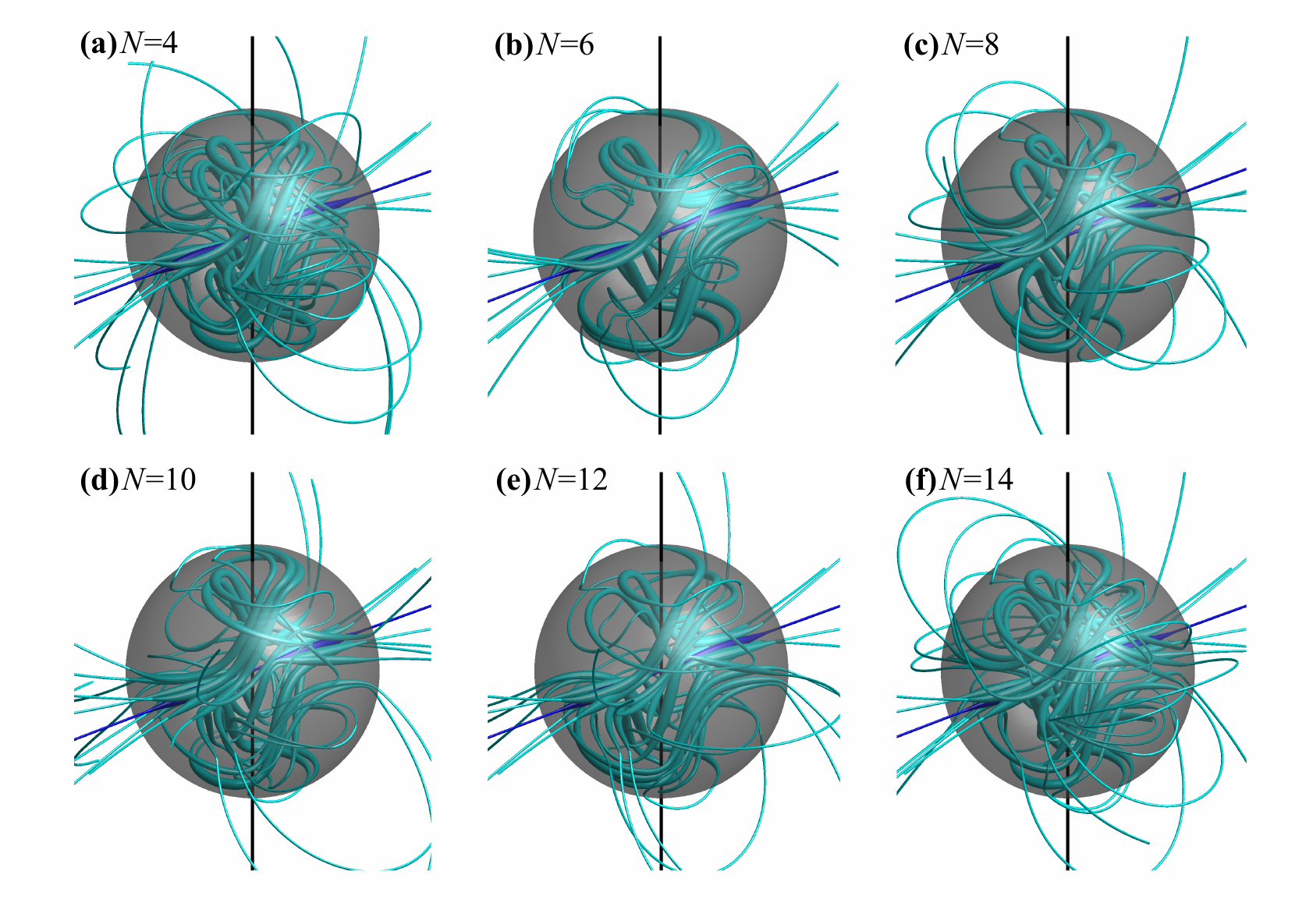}
\caption{Magnetic field lines of the fastest kinematic dynamo eigen-modes  obtained at $Rm=400$ for the optimized  flows shown in Fig.~\ref{v_tot}. Thickness of the lines is proportional to the magnitude of the field. Vertical lines represent the axis of symmetry of the flows. Horizontal lines (shown in darker color) denote the axis of these equatorial dipole-like dynamo fields. }\label{B_line_tot}
\end{figure}

Fig.~\ref{gam_tot} shows the dependences of dynamo growth rate $\gamma_b$ on magnetic Reynolds number $Rm$ obtained for the optimized flows from Fig.~\ref{v_tot}. A dynamo can be excited only in a relatively narrow range of $Rm$. In addition to $Rm_{cr}$ required for the dynamo onset, there is the second $Rm_{cr2}$, above which the dynamo is quenched. This is a typical indication of slow dynamos \cite{Childress_1995}.  The structures of the growing dynamo eigenmodes at $Rm=400$ are presented in Fig.~\ref{B_line_tot}. The excited dynamo field outside the sphere has dipole-like structure and the axis of dipole is perpendicular to the axis of flow symmetry.   

\section{Summary} 

We have numerically found the optimized laminar axisymmetric flows in a sphere that minimize the critical magnetic Reynolds number $Rm_{cr}$ required for the dynamo action. The flows are solutions to the Navier-Stokes equation with von K\'arm\'an type  boundary conditions specified by azimuthal velocity profiles $v_\phi(\theta)$ at the sphere's  wall. In this class of flows,  the overall minimum of $Rm_{cr}\approx200$ is obtained at the fluid Reynolds number $Re\approx240$ when the flow is hydrodynamically  marginally stable. Solving the kinematic dynamo problem for the optimized flows, we have determined that the dynamo action is quenched above the second critical magnetic Reynolds number $Rm_{cr2}$, which is typical for slow dynamos. 

We have shown that in a boundary driven flow  the dynamo can be excited only when the ratio of poloidal to toroidal kinetic energies in that flow is sufficiently high. This condition is necessary for the dynamo onset but does not guarantee it. In addition, the dynamo onset is very sensitive to the details of  driving velocity profile. In all cases explored here, axisymmetric flows of von K\'arm\'an type sustain dynamo only if the corresponding driving velocities have reversals (sign changes) near the equator. The effect of these reversals on dynamo onset is not fully comprehended.

Our results suggest that the dynamo excitation can be demonstrated in an experiment with controllable laminar plasma flows, such as MPDX. Simple estimates show that the dynamo regime with fluid Reynolds number of $Re=150$ and magnetic Reynolds number of $Re=400$ can be reached in an argon plasma with peak driving velocity of $V_0=5$~km/s, density $n_0=10^{18}$~m$^{-3}$, electron and ion temperatures of $T_e=10$ eV and $T_i=1$ eV, respectively. These parameters will soon be achievable in the MPDX.

\acknowledgements
 
This work is supported by the National Science Foundation and the United States Department of Energy.

\appendix
\section{Equations and numerical methods}
\label{app}

In a spherical harmonic basis the divergence-free vector fields $\v$, $\tilde\v$ and $\b$ are represented as \cite{Bullard_1954}
\begin{subequations}\label{v}
\begin{eqnarray}
\label{vr} v_r&=&\sum\limits_{l=1}^{L_0}\frac{l(l+1)s_l(r)Y_l^0}{r^2},\\
\label{vt} v_\theta&=&\sum\limits_{l=1}^{L_0}\frac{1}{r}\frac{\pa s_l(r)}{\pa r}\frac{\pa Y_l^0}{\pa\theta},\\
\label{vp} v_\phi&=&-\sum\limits_{l=1}^{L_0}\frac{t_l(r)}{r}\frac{\pa Y_l^0}{\pa\theta},
\end{eqnarray}     
\end{subequations}
\begin{subequations}\label{dv}
\begin{eqnarray}
\label{dvr} \tilde v_r&=&\sum\limits_{l=m}^{L_v}\frac{l(l+1)\tilde s_l(r)Y_l^m}{r^2},\\
\label{dvt} \tilde v_\theta&=&\sum\limits_{l=m}^{L_v}\bigg[\frac{1}{r}\frac{\pa \tilde s_l(r)}{\pa r}\frac{\pa Y_l^m}{\pa\theta}+\frac{im\tilde t_l(r)Y_l^m}{r\sin{\theta}}\bigg],\\
\label{dvp} \tilde v_\phi&=&\sum\limits_{l=m}^{L_v}\bigg[\frac{imY_l^m}{r\sin{\theta}}\frac{\pa \tilde s_l(r)}{\pa r}-\frac{\tilde t_l(r)}{r}\frac{\pa Y_l^m}{\pa\theta}\bigg],
\end{eqnarray} 
\end{subequations}
\begin{subequations}\label{B}
\begin{eqnarray}
\label{Br} B_r&=&\sum\limits_{l=m}^{L_b}\frac{l(l+1)S_l(r)Y_l^m}{r^2},\\
\label{Bt} B_\theta&=&\sum\limits_{l=m}^{L_b}\bigg[\frac{1}{r}\frac{\pa S_l(r)}{\pa r}\frac{\pa Y_l^m}{\pa\theta}+\frac{imT_l(r)Y_l^m}{r\sin{\theta}}\bigg],\\
\label{Bp} B_\phi&=&\sum\limits_{l=m}^{L_b}\bigg[\frac{imY_l^m}{r\sin{\theta}}\frac{\pa S_l(r)}{\pa r}-\frac{T_l(r)}{r}\frac{\pa Y_l^m}{\pa\theta}\bigg],
\end{eqnarray} 
\end{subequations}
where $Y_l^m$ are the spherical harmonics defined in terms of the associated Legendre polynomials as $Y_l^m(\theta,\phi)=P_l^m(\cos\theta)e^{im\phi}$. Since the equilibrium velocity field $\v$ is axisymmetric, and Eqs.~(\ref{mhd_v}), (\ref{mhd_b}) are linear in $\tilde\v$ and $\B$, each azimuthal mode $m$ of $\tilde\v$ and $\B$ can be considered separately. The summations in Eqs.~(\ref{v})-(\ref{B}) are truncated at some (generally different) spherical harmonics of degrees $L_0$, $L_v$ and $L_b$, respectively. In numerical calculations we discretize the unknown functions of radius $r$ ($0\leq r\leq1$) on  a uniform radial grid with $N_r$ intervals and apply the finite difference method of the second order. All results reported in the paper are obtained with  $L_0=L_v=L_b=20$ and  $N_r=50$. This spatial resolution appears to be sufficient for the cases under consideration. The convergence of  the numerical schemes is checked by comparing simulations at different resolutions. 

In the following subsections we consider in more details Eqs.~(\ref{mhd_NS})-(\ref{mhd_b}) for the fields given by Eqs.~(\ref{v})-(\ref{B}) and methods of their solution. 

\subsection{Equilibrium velocity}      

Substituting Eqs.~(\ref{v}) into Eq.~(\ref{mhd_NS}) and using the orthogonality properties of spherical harmonics, we obtain for $1\leq l\leq L_0$:
\begin{subequations}\label{st}
\begin{eqnarray}
\label{s}  \Delta_l^2s_l&=&Re\,A_l^0\sum\limits_{j=1}^{L_0}\sum\limits_{k=1}^{L_0}\bigg[ C_{ljk}\frac{\pa}{\pa r}\bigg(\frac{s_j\Delta_k s_k+ t_j t_k}{r^2}\bigg)
 + C_{jlk}\bigg(\frac{\Delta_k s_k}{r^2}\frac{\pa s_j}{\pa r} + \frac{t_k}{r^2}\frac{\pa t_j}{\pa r}\bigg) \bigg],\\ 
\label{t} \Delta_l t_l&=&Re\,A_l^0\sum\limits_{j=1}^{L_0}\sum\limits_{k=1}^{L_0}\frac{C_{ljk}}{r^2} \bigg[s_j\frac{\pa t_k}{\pa r} - t_j\frac{\pa s_k}{\pa r} \bigg].
\end{eqnarray} 
\end{subequations}
Here, $\Delta_l$ is the differential operator
$$
\Delta_l=\frac{\pa^2}{\pa r^2}-\frac{l(l+1)}{r^2},
$$
$A_l^m$ is the numerical factor (we give the general expression for arbitrary $m$, since it is used below)
$$
A_l^m=\frac{(2l+1)(l-m)!}{2l(l+1)(l+m)!},
$$
and elements  $C_{ljk}$ are defined as
$$
C_{ljk}=j(j+1)\int\limits_{0}^{\pi} Y_j^0\frac{\pa Y_l^0}{\pa\theta}\frac{\pa Y_k^0}{\pa\theta} \sin{\theta}d\theta.
$$
The boundary conditions for functions $s_l(r)$ and $t_l(r)$ in Eqs.~(\ref{st}) follow from the absence of a singularity in the velocity field at the center of the sphere and Eq.~(\ref{BCv}):
\begin{subequations}\label{BCst}
\begin{eqnarray}
\label{BCst0} s_l\big|_{r=0}&=&\frac{\pa s_l}{\pa r}\bigg|_{r=0}=0,~~~t_l\big|_{r=0}=0,\\
\label{BCst1} s_l\big|_{r=1}&=&\frac{\pa s_l}{\pa r}\bigg|_{r=1}=0,~~~t_l\big|_{r=1}=\tau_l.
\end{eqnarray}
\end{subequations}
Here $\tau_l$ are coefficients of the expansion of driving velocity $v_\phi(\theta)$ in terms of $(-\pa Y_l^0/\pa\theta)$:
\begin{equation}\label{vb}
v_\phi(\theta)=-\sum\limits_{l=1}^{L_0}\tau_l\frac{\pa Y_l^0}{\pa\theta}.
\end{equation} 
Coefficients $\tau_l$ are uniquely determined by Fourier coefficients $a_n$ from Eq.~(\ref{vb_VK}). Indeed, comparing Eqs.~(\ref{vb}) and (\ref{vb_VK}) we have
$$
-\sum\limits_{l=1}^{L_0}\tau_l\frac{\pa Y_l^0}{\pa\theta}=\sum\limits_{n=1}^{L_0}a_n\sin n\theta,
$$
where for generality we assumed Fourier expansion of $v_\phi(\theta)$, which includes both even and odd harmonics of $\theta$. Multiplying both sides of this equality by $(-A_l^0\sin\theta\,\pa Y_l^0/\pa\theta)$, integrating over $0\leq\theta\leq\pi$ and using orthogonality properties of the spherical harmonics we arrive at invertible matrix transform:
$$
\tau_l=\sum\limits_{n=1}^{L_0} F_{ln}a_n,~~~1\leq l,n\leq L_0 
$$   
with matrix elements $F_{ln}$ given by
$$
F_{ln}=-\frac{2l+1}{2l(l+1)}\int\limits_0^\pi\sin n\theta\sin\theta\frac{\pa Y_l^0}{\pa\theta}d\theta.
$$

In order to solve Eqs.~(\ref{st}) with boundary conditions given by Eqs.~(\ref{BCst}), we use an iterative scheme. The iterations are organized in the following way. First, by inverting the operators $\Delta_l^2$ and $\Delta_l$ (which are tridiagonal square matrices in finite difference representation) we bring Eqs.~(\ref{st}) to the form
$$
\x=\f(\x),~~~\x=(s_l,t_l), 
$$       
where $\f(\x)$ denotes the nonlinear right-hand side of Eqs.~(\ref{st}). Then we construct the iteration step:
\begin{equation}\label{iter}
\x^{(p+1)}=\x^{(p)}+\alpha\big[\f(\x^{(p)})-\x^{(p)}\big],
\end{equation}
where $\alpha$ is a constant, chosen to guarantee convergence of iterations. The iterations can be initialized with some profiles of $s^{(0)}_l(r)$ and $t^{(0)}_l(r)$  satisfying the boundary conditions. The iterations stop when 
$$
\|\x^{(p)}-\f(\x^{(p)})\|<\epsilon\|\x^{(p)}\|,
$$  
where $\epsilon$ is the error tolerance and the norm $\|\x\|$ is defined as a sum of squares of absolute values of all elements in $\x$. The results of the paper are obtained with $s^{(0)}_l(r)=0$, $t^{(0)}_l(r)=\tau_lr^{l+1}$, $\alpha=0.01$, $\epsilon=10^{-12}$. This choice provides fast convergence to equilibrium state with typical number of required iterations $\sim10^3$.         

\subsection{Hydrodynamic stability}   

Substituting Eqs.~(\ref{v}) and (\ref{dv}) into Eq.~(\ref{mhd_v}), we obtain for $m\leq l\leq L_v$:     
\begin{subequations}\label{dst}
\begin{eqnarray}
\label{ds}  \gamma_v\Delta_l\tilde s_l&=&\Delta_l^2\tilde s_l-ReA_l^m\sum\limits_{j=m}^{L_v}\bigg[\overline{I_{jl}^{(1)}}\Delta_j\tilde s_j + \overline{J_{jl}^{(4)}}\frac{\pa \tilde s_j}{\pa r}+\frac{\pa}{\pa r}\bigg(I_{lj}^{(2)}\Delta_j\tilde s_j + J_{lj}^{(3)}\frac{\pa \tilde s_j}{\pa r} + J_{lj}^{(4)}\tilde s_j \\ 
&-& I_{lj}^{(3)}\frac{\pa \tilde t_j}{\pa r} - I_{lj}^{(4)}\tilde t_j + J_{lj}^{(2)}\tilde t_j\bigg) - \overline{I_{jl}^{(4)}}\frac{\pa \tilde t_j}{\pa r} + \overline{J_{jl}^{(1)}}\tilde t_j  \bigg],\nonumber\\ 
\label{dt} \gamma_v\tilde t_l&=&\Delta_l \tilde t_l-Re\,A_l^m\sum\limits_{j=m}^{L_v}\bigg[I_{lj}^{(2)}\frac{\pa \tilde t_j}{\pa r} -I_{lj}^{(1)}\tilde t_j + J_{lj}^{(3)}\tilde t_j + I_{lj}^{(3)} \Delta_j\tilde s_j  - J_{lj}^{(2)}\frac{\pa \tilde s_j}{\pa r} + J_{lj}^{(1)}\tilde s_j \bigg].~~~~~~
\end{eqnarray} 
\end{subequations}
Here the bar above a symbol denotes its complex conjugate, $I_{lj}^{(1-4)}$ are functions of $r$ determined by equilibrium profiles $s_k(r)$, $t_k(r)$,
\begin{subequations}\label{I}
\begin{eqnarray}
\label{I1} I^{(1)}_{lj}[s_k,t_k]&=&\frac{j(j+1)}{r^2}\sum\limits_{k=1}^{L_v} \bigg[\frac{\pa s_k}{\pa r}M_{klj}+t_kL_{klj}\bigg],\\
\label{I2} I^{(2)}_{lj}[s_k,t_k]&=&\sum\limits_{k=1}^{L_v} \frac{k(k+1)s_k}{r^2}\bigg[l(l+1)K_{klj}-M_{klj}\bigg],\\
\label{I3} I^{(3)}_{lj}[s_k,t_k]&=&\sum\limits_{k=1}^{L_v} \frac{k(k+1)s_k}{r^2}\,L_{klj},\\
\label{I4} I^{(4)}_{lj}[s_k,t_k]&=&\frac{j(j+1)}{r^2}\sum\limits_{k=1}^{L_v}\bigg[\frac{\pa s_k}{\pa r}L_{klj} - t_k M_{klj}  \bigg],
\end{eqnarray} 
\end{subequations}
and $J_{lj}^{(1-4)}$ are obtained from Eqs.~(\ref{I}) by replacement  $s_k\to t_k$ and $t_k\to -\Delta_k s_k$, i.e., 
$$
J_{lj}^{(1-4)}=I_{lj}^{(1-4)}[t_k, -\Delta_ks_k].
$$
The elements $K_{klj}$, $L_{klj}$ and $M_{klj}$ in Eqs.~(\ref{I}) are integrals of triple products of spherical harmonics:  
\begin{eqnarray}
\label{K} K_{klj}&=&\int\limits_{0}^{\pi}  Y_k^0 Y_l^m\overline{Y_j^m} \sin{\theta}d\theta,\nonumber\\
\label{L} L_{klj}&=&im\int\limits_{0}^{\pi}  \frac{\pa Y_k^0}{\pa\theta} Y_l^m\overline{Y_j^m} d\theta,\nonumber\\
\label{M} M_{klj}&=&\int\limits_{0}^{\pi}  \frac{\pa Y_k^0}{\pa\theta}\frac{\pa Y_l^m}{\pa\theta}\overline{Y_j^m} \sin{\theta}d\theta.\nonumber
\end{eqnarray} 
The indices in the above formulas span the following intervals: $1 \leq k\leq L_0$, $m \leq l, j\leq L_v$. Note that $M_{klj}$ can be reduced to $K_{klj}$ by integration by parts, namely,
$$
M_{klj}=\frac{1}{2}\big[k(k+1)+l(l+1)-j(j+1)\big]K_{klj}.
$$
The boundary conditions for functions $\tilde s_l(r)$ and $\tilde t_l(r)$ are 
\begin{subequations}\label{BCdst}
\begin{eqnarray}
\label{BCdst0} \tilde s_l\big|_{r=0}&=&\frac{\pa \tilde s_l}{\pa r}\bigg|_{r=0}=0,~~~\tilde t_l\big|_{r=0}=0,\\
\label{BCdst1} \tilde s_l\big|_{r=1}&=&\frac{\pa \tilde s_l}{\pa r}\bigg|_{r=1}=0,~~~\tilde t_l\big|_{r=1}=0.
\end{eqnarray}
\end{subequations}

Eqs.~(\ref{dst}) and (\ref{BCdst}) constitute an eigenvalue problem for the growth rate $\gamma_v$ and unknown functions $\tilde s_l(r)$ and $\tilde t_l(r)$ describing the velocity perturbations for the azimuthal mode $m$. Applying the finite difference method, we reduce the problem to a generalized matrix eigenvalue equation with matrix size $[2(L_v-m+1)(N_r-1)]^2$. This equation is solved in MATLAB.

\subsection{Kinematic dynamo}  

Substituting Eqs.~(\ref{v}) and (\ref{B}) into Eq.~(\ref{mhd_b}) we obtain for $m\leq l\leq L_b$:     
\begin{subequations}\label{ST}
\begin{eqnarray}
\label{S} \gamma_b S_l&=&\Delta_l S_l+Rm\,A_l^m\sum\limits_{j=m}^{L_b}\bigg[I^{(1)}_{lj}S_j 
-I^{(2)}_{lj}\frac{\pa S_j}{\pa r} + I^{(3)}_{lj}T_j \bigg],\\
\label{T} \gamma_b T_l&=&\Delta_l T_l-Rm\,A_l^m\sum\limits_{j=m}^{L_b}\bigg[\overline{I^{(1)}_{jl}}T_j + \frac{\pa}{\pa r}\bigg(I^{(2)}_{lj}T_j+I^{(3)}_{lj}\frac{\pa S_j}{\pa r} + I^{(4)}_{lj}S_j\bigg)+ \overline{I^{(4)}_{jl}}\frac{\pa S_j}{\pa r} \bigg],~~~
\end{eqnarray} 
\end{subequations}
where integrals $I_{lj}^{(1-4)}$ are defined by Eqs.~(\ref{I}) with indices $l,j$ spanning the interval $m\leq l,j\leq L_b$. The boundary conditions for the non-ferritic insulating wall are obtained by matching the dynamo  field onto a vacuum potential field solution at $r=1$: 
\begin{subequations}\label{BCST}
\begin{eqnarray}
\label{BCST0} S_l\big|_{r=0}&=&0,~~~T_l\big|_{r=0}=0,\\
\label{BCST1} \left(\frac{\pa S_l}{\pa r} + lS_l \right)\bigg|_{r=1}&=&0,~~~ T_l\big|_{r=1}=0.
\end{eqnarray}    
\end{subequations}
Formulas analogous to Eqs.~(\ref{ST}), (\ref{BCST}) are originally derived in Ref.~\cite{Bullard_1954}.

We transform Eqs.~(\ref{ST}) and (\ref{BCST}) to a matrix eigenvalue problem by applying finite difference method. In this method, the boundary condition for $S_l$ in  Eqs.~(\ref{BCST1}) is taken into account by using an extra (ghost) grid point to approximate the derivative at $r=1$. The resulting matrix equation of size $[(L_b-m+1)(2N_r-1)]^2$ is solved in MATLAB.

\end{document}